\documentclass[prd,twocolumn,nofootinbib,showpacs,eqsecnum,floatfix,superscriptaddress]{revtex4}

\usepackage{amsmath}
\usepackage{graphicx}
\usepackage{bbm}
\usepackage{color}
\usepackage{bm}
\usepackage{latexsym}
\usepackage{amssymb}
\usepackage{booktabs}
\usepackage[ansinew]{inputenc}
\usepackage{rotating}
\usepackage{euscript}
\usepackage{gensymb}

\newcommand{\eq}[1]{\begin{equation}\label{#1}}
\newcommand{\en}{\end{equation}}

\newcommand{\bmath}[1]{\ensuremath{\bm{#1}}}

\newcommand{\SU}{\mathrm{SU}}

\newcommand{\be}{\begin{equation}}

\begin{document}

\title{Fluctuations of the baryonic flux-tube
junction from effective string theory}

\author{Melanie~Pfeuffer} 
\affiliation{Institute for Theoretical Physics, University of
Regensburg, 93040 Regensburg, Germany}

\author{Gunnar~S.~Bali} 
\affiliation{Institute for Theoretical Physics, University of
Regensburg, 93040 Regensburg, Germany}

\author{Marco~Panero} 
\affiliation{Institute for Theoretical Physics, University of
Regensburg, 93040 Regensburg, Germany} 
\affiliation{Institute for Theoretical Physics, ETH Z\"urich, 8093 Z\"urich, Switzerland}

\date{\today}

\begin{abstract}
 In quenched QCD, where the dynamic creation of quark-antiquark pairs out of the vacuum is neglected, a confined baryonic system composed of three static quarks exhibits stringlike behavior at large interquark separation, with the formation of flux tubes characterized by the geometry of the so-called \emph{Y} ansatz. We study the fluctuations of the junction of the three flux tubes, assuming the dynamics to be governed by an effective bosonic string model. We show that the asymptotic behavior of the effective width of the junction grows logarithmically with the distance between the sources, with the coefficient depending on the number of joining strings, on the dimension of spacetime and on the string tension.
\end{abstract}


\pacs{12.38.Aw, 11.25.-w, 21.30.Fe
}

\maketitle 

\section{Introduction}
\label{sec:intro}
One of the crucial features characterizing quantum chromodynamics
(QCD) is confinement: the fundamental constituents of strongly
interacting matter (quarks and gluons) are not observed as asymptotic
states, and the low-energy hadronic spectrum consists of colorless states
only.

Because of its nonperturbative nature, a formal proof of confinement from
first principles has so far been elusive, and even
the degrees of freedom responsible for this phenomenon are subject to debate.
However, some
low-energy properties of hadrons and QCD forces --- see, e.g.\
Ref.~\cite{Bali:2000gf} for a review --- can be accurately modeled in
terms of an effective string picture~\cite{Luscher:1980fr, Luscher:1980ac, Luscher:1980iy,Arvis:1983fp,Luscher:2002qv},
which describes the infrared
properties of hadrons in terms of a fluctuating, thin (almost unidimensional)
flux tube joining the color sources. At
sufficiently large interquark separations the lowest-energy excitations
of the confined system are associated with collective degrees of freedom
corresponding to transverse stringlike vibrations of the flux tube, whereas the excitation spectrum of the gauge degrees of freedom inside
the tube is much higher lying.

While it is unproven that low-energy aspects of
confining quantum field theories can indeed be explained in terms
of an effective string theory with universal features,
strong evidence in favor of this conjecture is provided by
recent comparisons between results from lattice simulations and bosonic
string predictions (surveyed below).
The observation of a linearly
rising potential between color sources separated at distances
$R$ does not automatically
imply the existence of an effective string description,
however, the universality of the subleading $1/R$-coefficient found in
several gauge models~\cite{deForcrand:1984cz,Bali:1992ab, Luscher:2002qv, Juge:2002br,Bali:1994de,  Majumdar:2002mr, Caselle:2002ah, Caselle:2004er, Caselle:2007yc, Panero:2005iu, Lucini:2001nv, Gliozzi:2005ny, Athenodorou:2007du} provides such a nontrivial test.
At present not many
quantitative predictions that exceed the classical limit are
available to compare lattice data to.
In this article we increase the number of such
nontrivial predictions by
calculating the width of ``baryonic'' flux-tube junctions for
general geometries.

Note that while we label our configurations as baryonic,
we expect our predictions only to apply to
pure Yang-Mills theories with static external charges.
The bosonic string model is unlikely to be a good approximation
to the baryons of real QCD with sea quarks, which are likely
to decay
into baryon-meson pairs, before the string limit of large
distances can be reached. 

For the simplest physical scenario, i.e.\ a (``mesonlike'') pair of static,
infinitely heavy, confined color sources, the effective model has been
developed since the early 1980s~\cite{Luscher:1980fr, Luscher:1980ac, Luscher:1980iy,Arvis:1983fp}.
Later on this description was reformulated in terms of an expansion about
the long-string vacuum~\cite{Polchinski:1991ax}. More recent
theoretical developments include
Refs.~\cite{Luscher:2004ib, Drummond:2004yp, Hari_Dass:2007gn},
and are reviewed in Ref.~\cite{Kuti:2005xg}.
The large-distance string behavior has been observed in numerical lattice
simulations of the torelon spectrum (corresponding to closed strings)
and static potentials (corresponding to mesonic open strings)
of $\SU(3)$ lattice gauge
theory~\cite{deForcrand:1984cz,Bali:1992ab, Luscher:2002qv, Juge:2002br} as well as of
various other gauge models~\cite{Bali:1994de,  Majumdar:2002mr, Caselle:2002ah, Caselle:2004er, Caselle:2007yc, Panero:2005iu, Lucini:2001nv, Gliozzi:2005ny, Athenodorou:2007du}.

The effective string picture also predicts the
width of the flux tube to grow logarithmically
as a function of the interquark distance as well as the
coefficient of the logarithm~\cite{Luscher:1980iy};
this was addressed and confirmed in various numerical
lattice studies~\cite{Bali:1994de,Pennanen:1997qm,Boyko:2007ae,Caselle:1995fh, Zach:1997yz, Koma:2003gi, Chernodub:2007wi, Giudice:2006hw}.

\begin{figure*}
 \includegraphics[width=.98\textwidth]{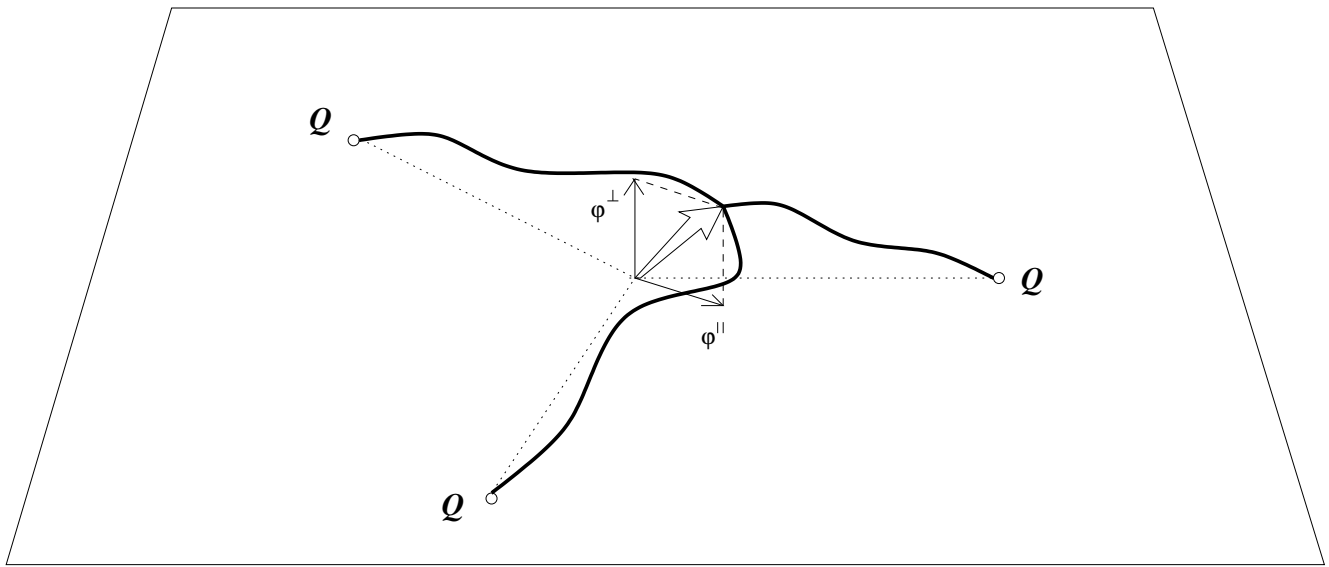}\caption{A snapshot of the fluctuating flux tubes (idealized as strings) of three static color sources $Q$, joined at a common junction; $\bmath{ \varphi}$ denotes the position of this junction, relative to its classical location minimizing the total string length. The components of $\bmath{ \varphi}$ parallel and normal to the plane containing the static color sources are also displayed.}\label{fig:fluctuating_strings}
\end{figure*}

Lattice simulations of the baryonic
setup~\cite{Bali:2000gf,Alexandrou:2001ip,Alexandrou:2002sn,Takahashi:2002bw,
deForcrand:2005vv}
indicate that the flux-tube
profile interpolates between the so-called $\Delta$ geometry
at short distances (where the effective one-gluon exchange dominates),
and the $Y$ ansatz for separations between the sources of the order
of or larger than approximately 0.8~fm. This $Y$ ansatz
that is relevant in the infrared region is
characterized by a junction where the flux tubes
meet (see Fig.~\ref{fig:fluctuating_strings}).
The corresponding leading order string corrections to
the baryonic potential have been worked out in Ref.~\cite{Jahn:2003uz}.

Building upon the procedure used in this reference, in this article we
study the width of the junction, assuming that this is generated
by string fluctuations
of an effective string theory with the lowest dimensional
term given by the Nambu-Goto (NG) action~\cite{Nambu:1974zg, Goto:1971ce},
with string tension $\sigma$. We consider
the leading order nontrivial behavior in the limit
of large separations between the static color sources.

\section{Calculation setup}
\label{sec:setup}

The calculation is performed in $D$-dimensional Euclidean spacetime
with $D-1$ spatial dimensions of infinite extent and
a periodic time coordinate $t\in[0,T)$.
We consider the general case of $2\leq n\leq D$ static
``quarks'' spanning a $(n-1)$-dimensional hyperplane. We assume the
action to be minimal when the $n$ strings meet in a common junction.
For $n=3$ this is indeed the case, unless one of the angles of the
triangle defined by the three sources exceeds the critical
value $2\pi/3$.
In this latter case the junction
will be fixed to the position of the corresponding source and the system
can be decomposed into two mesonic strings.
Note that also
for many $n>3$ geometries the classical configuration
will be characterized by different geometries, with two or more distinct
junctions (see also Ref.~\cite{Gliozzi:2005en}). In principle
our calculation can be extended to these cases.

\begin{figure*}
\includegraphics[width=.41\textwidth]{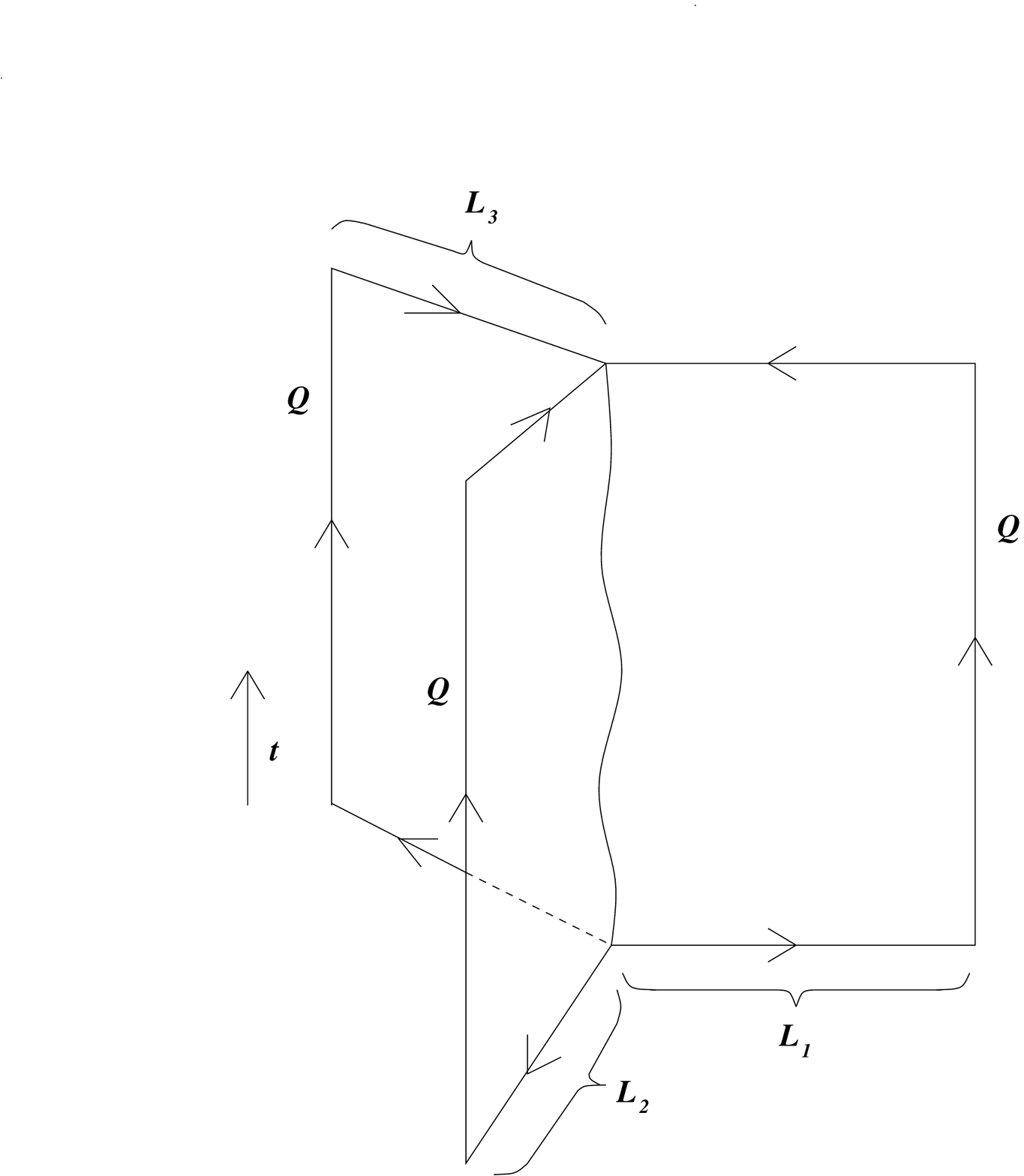}\hfill
\includegraphics[width=.58\textwidth]{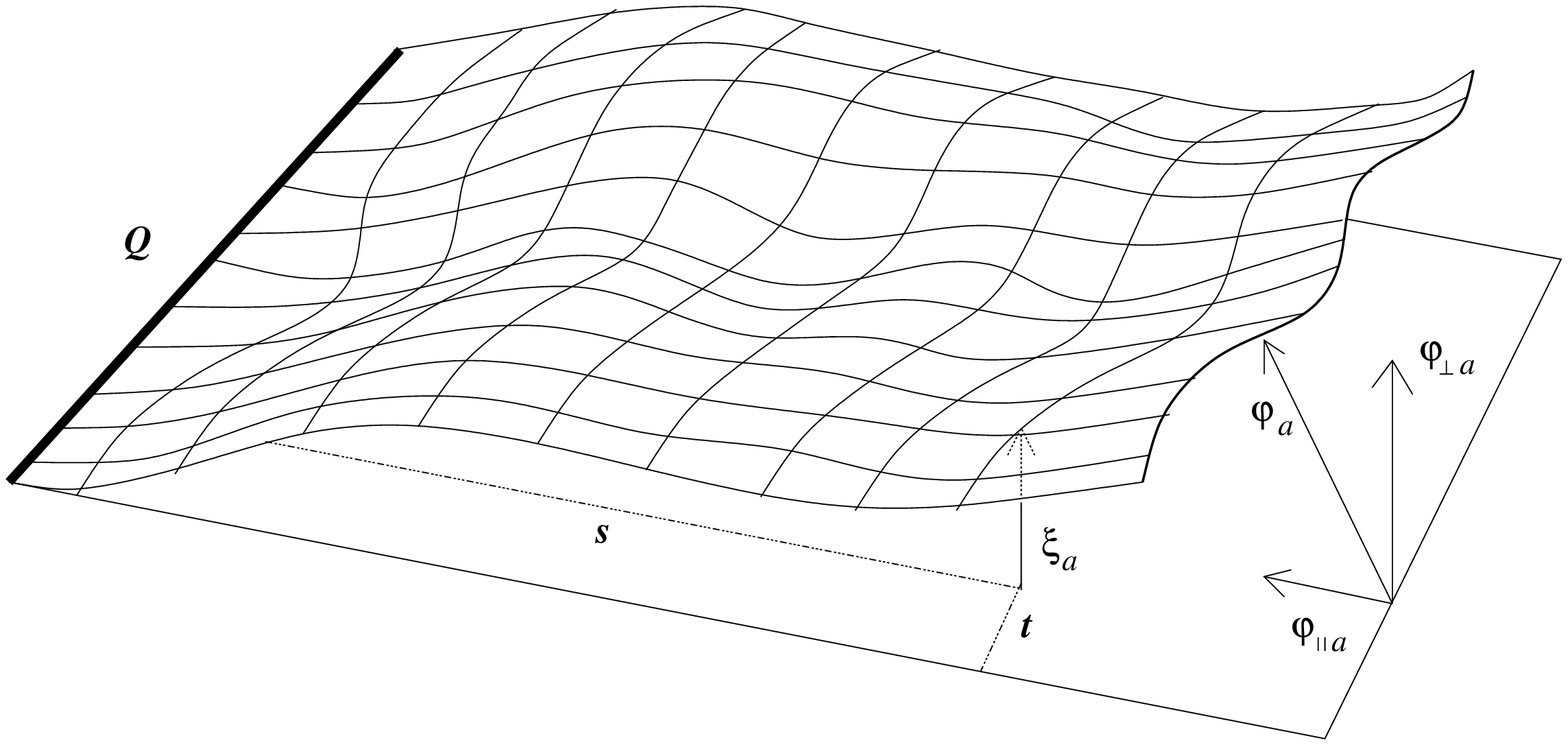}
  \caption{Left-hand side: World sheets spanned by the fluctuating strings
during their time evolution. Right-hand side: Surface of one of the
string world sheets.}
  \label{fig:blades_and_one_blade}
\end{figure*}

During their time evolution, the strings
span $n$ different world sheets
(see Fig.~\ref{fig:blades_and_one_blade}); each of these blades is bounded by
the (straight) worldline of a static quark on one side, and a
generic worldline spanned by the fluctuating junction on the other side.

Classically, the ground state fulfills the constraint of the minimal area
of the string world sheets.
Therefore the position of the junction is determined by the requirement
of minimal total string length. Furthermore, the balance of tensions
means that $\sigma \sum_{a=1}^n \bmath{ e}_a =\bmath{ 0}$
which implies equal angles between the
strings at any time.
$\bmath{ e}_a$ denote unit vectors along the direction between
the junction and the quark $a$.
Assuming that the string dynamics is described by the NG
action $S_{\rm NG}$ means that the quantum weight of a generic
configuration of a string world sheet is proportional to
the exponential of its total area. In formul\ae:
\begin{equation}
  S_{\rm NG}=\sigma \int\! d^2 \!{\rm\zeta}\, \sqrt{- g}\:,
\label{eq:NG}
\end{equation}
where the position on the surface of the string world sheet is
parametrized by the coordinates ${\rm \zeta}_i, i\in\{1,2\}$.
Although this bosonic string model is nonrenormalizable
(because it is nonpolynomial) and anomalous (except in $26$ spacetime
dimensions)~\cite{Goddard:1973qh,Brink:1973ja,Arvis:1983fp},
it can be considered a legitimate
starting point in the construction of an effective theory, and
it can be shown to agree with the Polchinski-Strominger
effective model~\cite{Polchinski:1991ax} up to and including
the next-to-leading order~\cite{Drummond:2004yp}.
Let ${\rm X}^\mu({\bm \zeta})$, with $\mu\in\{1,\ldots,D\}$, denote a map from the world sheet to the spacetime,
embedding the world sheet; the induced metric
of Eq.~(\ref{eq:NG}) is given by
\begin{equation}
  g^{ij}=\frac{\partial {\rm X}^\mu}{\partial {\rm\zeta}_i}\frac{\partial{\rm
  X}_\mu}{\partial{\rm\zeta}_j}\:.
\end{equation}
To proceed with the quantum treatment, it is natural to fix the
reparametrization and Weyl invariance to the ``physical'' gauge,
allowing us to describe the transverse
displacements (for our purposes, the Weyl anomaly can be neglected, because it vanishes in the limit of large distances). This means that
only transverse fluctuations $\bmath{\xi}_a (t,s)$ of the string
world sheets $a\in\{1,\ldots,n\}$
around the classical configuration are considered as 
physical. The time $t$ and parameter $s$ label the position on
string world sheet (blade) $a$. In particular, $s$ denotes the
spatial distance from
the worldline of the quark $a$, i.e.\ the classical
position of the junction is given by $s=L_a$. The junction
worldline $\bmath{ \varphi}(t)$ fluctuates within the hyperplane spanned
by the quarks (changing the minimal area of the blades) as well as in the
$D-n$ remaining transverse spatial directions.
From continuity,
we obtain the boundary conditions for the transverse fluctuations
$\bmath{\xi}_a (t,s)$:
\be
\label{boundarycondition_xi}
\bmath{\xi}_a(t,L_a+\bmath{ e}_a \cdot \bmath{ \varphi}(t)) = \bmath{ \varphi}_{\perp a}(t)\:,
\en
where, $\bmath{ \varphi}_{\perp a}\equiv\bmath{ \varphi}
-\bmath{ e}_a({\bmath{ e}}_a \cdot
\bmath{ \varphi})$. The transverse fluctuations $\bmath{\xi}_a (t,s)$
vanish at the location of the quarks ($s=0$), and are periodic in the
time $t$, with period $T$.

For technical reasons we assume the
junction itself to have a finite mass $m$.
This results
in a static energy and in a kinetic term.
The parameter value $m$ should not affect the large-distance
results, $L_a\sigma\gg m$, that we present below, and indeed
it cancels in the calculation.
Expanding the NG action around the equilibrium configuration yields
\begin{equation}\label{generalaction}
  S=S_\parallel+\frac{\sigma}{2} \sum_{a,i} \int_{\Gamma_a}
\!\!d^2\!{\rm\zeta}\,\, \frac{\partial
  \bmath{ \xi}_a}{\partial{\rm \zeta}_i} \cdot
\frac{\partial \bmath{\xi}_a}{\partial{\rm \zeta}_i} + m \left( T + \frac {1}{2}
  \int_0^T\! dt\, |\dot{\bmath{ \varphi}}|^2 \right),
\end{equation}
where again ${\rm \zeta}_1, {\rm \zeta}_2$ are world sheet parameters
and
\begin{equation}
  S_\parallel=\sigma \sum_a \left( L_a T + \int\! dt\,
  \bmath{ e}_a \cdot \bmath{ \varphi} (t) \right) = \sigma L_Y T\:.
\end{equation}
$L_Y = \sum_a L_a$ above denotes the
total string length. (Note that $\sum_a\bmath{ e}_a=\bmath{ 0}$.)
In the $T \rightarrow \infty$ limit
the string thickness can be extracted from the partition function, which
to leading nontrivial order is given by
\be
\label{total_partitionfct}
 Z = e^{-(\sigma L_Y +m)T}\int\!\EuScript{D} \bmath{ \varphi}~
 \exp\left(-\frac{m}{2}\int\!dt\,
 |\dot{\bmath{ \varphi}}|^2\right)\prod_{a=1}^3Z_a(\bmath{ \varphi})\:,
\en
where $Z_a(\bmath{ \varphi})$ denotes the partition function for the fluctuations of a given blade that is bounded by the
junction worldline ${\bmath{ \varphi}}(t)$:
\be
Z_a(\bmath{ \varphi})=
\int\!{\EuScript D}\bmath{\xi}_a\,\exp\left(-\frac{\sigma}{2}\int |\partial
 \bmath{\xi}_a|^2\right)\:.
\en
The string partition functions $Z_a(\bmath{ \varphi})$ are Gaussian
functional integrals and can be calculated as follows:
\be
\label{partitionfct_string}
  Z_a(\bmath{ \varphi})=e^{-\frac{\sigma}{2}\int |\partial
 \bmath{\xi}_{\min,a}|^2}
 |\det(-\triangle_{\Gamma_a})|^{-(D-2)/2}\:,
\en
where $\bmath{\xi}_{\min,a}$ is the minimal-area solution for given
$\bmath{ \varphi}(t)$. $\triangle_{\Gamma_a}$ denotes the Laplacian acting on
the domain (blade) $\Gamma_a$.
$\bmath{\xi}_{\min,a}(t,s)$ is harmonic and
satisfies the boundary conditions
Eq.~(\ref{boundarycondition_xi})~\cite{Jahn:2003uz}.
Below we will evaluate this expression to the leading order in terms of the
fluctuations $\bmath{ \varphi}$.

In contrast to the
mesonic setup, the world sheets $\Gamma_a$ are in general no rectangles.
However, the determinant in
Eq.~(\ref{partitionfct_string}) can still be
calculated by decomposing the boundary $\bmath{ \varphi}(t)$
of $\Gamma_a$ into
a sum over Fourier modes and
conformally mapping the resulting domains to rectangles, as shown in
Appendix~\ref{AppA}.
Carrying out this mapping and taking the limit $T\rightarrow\infty$,
Jahn and de~Forcrand~\cite{Jahn:2003uz}
derived the subleading term of the ($n=3$)
baryonic potential ${\rm V}_{qqq}$:
\begin{widetext}
\begin{eqnarray}
\label{vba1}
  {\rm V}_{qqq}(L_1,L_2,L_3)&=& \sigma\sum_aL_a + V^\parallel + (D-3)V^\perp +
                    {\EuScript O}(L^{-2}_a)\:,\nonumber\\\label{total_potential}
  V^\parallel &=&  -\frac{\pi}{24}\sum_a \frac{1}{L_a}+ \int_0^\infty \frac {d w}{2\pi}\ln\left[\frac 1 3 \sum_{a<b} \coth(w L_a)\coth(w L_b)\right]\:,\\
  V^\perp &=& -\frac{\pi}{24}\sum_a \frac{1}{L_a}+ \int_0^\infty \frac {d w}{2\pi}\ln\left[ \frac 1 3 \sum_a
                    \coth(w L_a)\right]\:.\nonumber
\end{eqnarray}
\end{widetext}
We confirm this result. For the equilateral case $L=L_1=L_2=L_3$
Eq.~(\ref{total_potential}) simplifies to:
\be
\label{vba2}
V_{qqq,\triangle}(L)=3\sigma L-\frac{D-3}{16}\frac{\pi}{L}+{\EuScript O}(L^{-2})\:.
\en

\section{String thickness at the junction}
The bosonic string model yields a prediction for the
thickness of the fluctuating strings. The width of the junction itself can be
calculated by taking the expectation value:
\begin{equation}
  \langle{\bmath{ \varphi}}^2\rangle
=\frac{\int\! {\EuScript D }\bmath{ \varphi}\,{\bmath{ \varphi}}^2
  e^{-S}}{\int\! {\EuScript D} \bmath{ \varphi}\, e^{-S}}\:.
\end{equation}
The action $S$ is defined in Eq.~(\ref{generalaction})
and can also be read off from the partition function
Eq.~(\ref{total_partitionfct}). We split the
string width into contributions $\langle{\bmath{ \varphi}}^{\perp
2}\rangle$, perpendicular to the hyperplane spanned by the $n$ quarks,
and $\langle{\bmath{ \varphi}}^{\parallel 2}\rangle$ within
the hyperplane of the quarks:
\begin{equation}
  \langle{\bmath{ \varphi}}^2\rangle=
\langle{\bmath{ \varphi}}^{\perp 2}\rangle+
\langle{\bmath{ \varphi}}^{\parallel 2}\rangle\:.
\end{equation}
Note that ${\bmath{ \varphi}}^{\perp}$ lives within a $(D-n)$-dimensional subspace and ${\bmath{ \varphi}}^{\parallel}$ fluctuates in
the remaining $n-1$ spatial directions. This differs from the
definitions with respect to a given blade $a$ of
Sec.~\ref{sec:setup} above,
${\bmath{ \varphi}}_{\perp a}$ [$(D-2)$-dimensional fluctuations]
and ${\bmath{ \varphi}}_{\parallel a}$ (one-dimensional fluctuations).

In the $T\rightarrow\infty$ limit, the perpendicular contribution
reads
\begin{equation}
  \langle{\bmath{ \varphi}}^{\perp 2}\rangle
  =  (D-n)\frac 1 \pi \int_0^\infty\! {dw}\,\frac{1}{mw^2+{\sigma} w \sum_a\coth(wL_a)}\:.\label{iperp}
\end{equation}
We present more details of the calculation in Appendix~\ref{Appendix_mu2}.
For $n$ strings of identical length $L=L_1 = L_2 = \cdots=L_n$ we obtain
\begin{equation}\label{eq_phi_perp}
  \langle{\bmath{ \varphi}}^{\perp 2}\rangle
  =  (D-n)\frac 1 \pi \int_0^\infty\!
{dw}\,\frac{1}{mw^2+{n \sigma} w \coth(wL)}\:.
\end{equation}
We split this integral at $w=C/L$, where $C$ is an arbitrary $L$-independent
constant. The first part of integration is subleading in $L$.
We can choose $C$ large enough, such that the
approximation $\coth(C)\approx 1$ holds.  Neglecting subleading terms
in $L$,
the second part of
integration gives
\begin{equation}
 \int_{C/L}^\infty\! {dw}\,\frac{1}{mw^2+{n \sigma} w \coth(wL)}
\simeq \frac 1 {n\sigma} \ln L\:.
\end{equation}
Therefore, to leading order, the result reads
\begin{equation}\label{mu_perp}
  \langle{\bmath{ \varphi}}^{\perp 2}\rangle=
  \frac{D-n} n \frac 1 {\pi\sigma}\ln\frac{L}{L_0}\:,
\end{equation}
where we have absorbed an arbitrary constant into $L_0$;
the width of the junction, orthogonal to the
plane spanned by the quarks, grows logarithmically with the distance.
$L_0$ will depend on $D$, $n$, and the microscopic (ultraviolet)
details of the gauge model.

We can perform a consistency check by comparing the above result
to the mesonic case ($n=2$).
The logarithmic behavior of the width of a flux tube connecting two
quarks in the string picture was predicted many years ago by
L\"{u}scher, M\"{u}nster and Weisz~\cite{Luscher:1980iy}.
The result they found for $D=4$ is
\begin{equation}
  \delta^2\sim\frac 1 {\pi M^2}\ln \frac{L'}{\lambda}\:,
\end{equation}
where $M^2$ denotes the string tension and $\lambda$ represents a cutoff
scale.
The effective string width was studied again by Caselle
\emph{et al.}~\cite{Caselle:1995fh} and by
Gliozzi~\cite{Gliozzi_1994},
calculating the deviation of the
transverse coordinates of the string from the respective
Green function. The result they obtained
for the mean squared width $w_0^2$ in $D=3$, determined at the symmetry point of the string world sheet is
\begin{equation}
  w_0^2=\frac 1 {2\pi\sigma}\ln\frac{R}{R_c}\:,
\end{equation}
where $\sigma$ denotes the string tension, $R$ the interquark
distance and $R_c$ is an ultraviolet scale.

We divide the string connecting quark and antiquark
into two parts of equal length (up to small longitudinal
fluctuations), connected in the middle
by a junction. We can then apply Eq.~(\ref{mu_perp})
for $D=3$ and $D=4$. The above predictions indeed
coincide with our results
where we identify
$\sigma=M^2$, $L=2L'=2R$ and $L_0=2\lambda=2R_c$.

Now let us turn to the string width within the plane of the quarks.
This is only well defined for $n\geq 3$ and
the most interesting case is the $n=3$ baryon.
In contrast to
the perpendicular width $\langle{\bmath{ \varphi}}^{\perp 2}\rangle$,
our calculation applies to $n=3$ only since we assume the sources
to lie in a two-dimensional plane. In this case the $n>3$ minimal
string configuration will usually contain more than one junction, unless
junctions are fixed at the positions of quarks and do not fluctuate.
It turns out that our calculation of 
$\langle{\bmath{ \varphi}}^{\parallel 2}\rangle$ cannot
easily be generalized to higher dimensional planes, i.e.\ to
$n>3$.
In the baryonic equilateral case we obtain [Eq.~(\ref{mupara_bary})]:
\begin{equation}\label{eq_phi_parallel}
 \langle{\bmath{ \varphi}}^{\parallel 2}_{qqq,\triangle}\rangle  =
       \frac {4}{3\sigma\pi}
                \int_0^\infty 
                 \frac {dw}
                        {w^2\tilde{m}/L + (w- w^3 a/L^2) \coth(w) }
\:,
\end{equation}
where $\tilde{m}=2m/(3\sigma)$ and $a=(D-2)/(12\pi\sigma)$.
The result for nonequilateral configurations can be obtained
{}from Eq.~(\ref{ipar}) below.

We split the integral in analogy to the above discussion of the
perpendicular fluctuations. However, one finds that,
like in the calculation of the baryonic potential, a pole emerges.
If one views the NG action as the first term within an effective
string theory then this pole has to be canceled by counterterms
arising from the inclusion of higher dimensional operators.
In this sense it should not affect the leading order
result. Assuming this,
the parallel contribution to the width of the junction turns out to be:
\begin{equation}
\label{iii}
  \langle{\bmath{ \varphi}}^{\parallel 2}_{qqq}\rangle =\frac 4 3
\frac {1}{\pi\sigma}\ln\frac{L}{L_c}\:,
\end{equation}
where again subleading contributions are suppressed and
$L_c$ is an undetermined constant.
Note that the coefficient above is by a factor
$4/(D-3)$ larger than the one in front of the logarithm within
the expression for
$\langle{\bmath{ \varphi}}^{\perp 2}\rangle$
of Eq.~(\ref{mu_perp}). One factor $2/(D-3)$ corresponds to
the ratio of independently fluctuating parallel over
perpendicular components while another factor of 2 is expected
from the stronger restoring force for perpendicular displacements,
relative to parallel ones.

\section{Conclusions}
In this article, we studied the width of the junction of
flux tubes in baryonlike systems composed of
infinitely heavy, static color sources (quarks) at
large distances $L$ from the junction.
Assuming the low-energy aspects to be governed by the
dynamics of the bosonic Nambu-Goto string model,
we have shown that the width of the junction grows
logarithmically with the distance between the quarks. In particular the
quadratic width orthogonal to the $(n-1)$-dimensional plane
spanned by $n$ equidistant quarks [the baryons of
$\SU(n)$ gauge theories] in $D\geq n$ spacetime dimensions 
reads
[Eq.~(\ref{mu_perp})]:
\be
  \langle{\bmath{ \varphi}}^{\perp 2}\rangle=
  \frac{D-n} n \frac 1 {\pi\sigma}\ln\frac{L}{L_0}\:.
\en
This also applies to (and generalizes) the mesonic case ($n=2$).
The corresponding result for
general geometries with a Steiner junction can be obtained
from Eq.~(\ref{iperp}). The width within the plane of the sources
also grows logarithmically as a function of the separation
and we have calculated this
for $n=3$. The result for the equilateral case
is displayed in Eq.~(\ref{iii}) while the general result
can be calculated from Eq.~(\ref{ipar}). We also confirm the
result of Ref.~\cite{Jahn:2003uz} for the baryonic
potential, Eqs.~(\ref{vba1}) -- (\ref{vba2}).

The mesonic flux-tube width has already
been investigated in lattice simulations
of different gauge
theories~\cite{Bali:1994de,Caselle:1995fh,Pennanen:1997qm,Zach:1997yz,Koma:2003gi,Giudice:2006hw,Chernodub:2007wi,Boyko:2007ae}. While most of
these studies confirm the logarithmic broadening of the mesonic string,
without much lattice spacing dependence, it
should be noted that Ref.~\cite{Boyko:2007ae} found such
broadening only at fixed lattice spacings but the string width
to actually shrink, possibly to zero, if the continuum limit was taken.
This is also incompatible with the result
of Ref.~\cite{Heinzl:2008tv} of a vanishing overlap between 
a thin string state and the ground
state wave function.
Further lattice studies are required to resolve this controversy.

The question if and at what distances
our string predictions become valid can be addressed by lattice simulations of
baryonic configurations in $\SU(3)$ gauge theory at large $L$
in $D=3$ and $D=4$ spacetime dimensions.
While this is numerically quite challenging, at least the
simplified case of $D=3$ $Z_3$ gauge theory can be mapped
to a two-dimensional Potts model, allowing for precise
numerical simulations~\cite{deForcrand:2005vv,Caselle:2005sf}.
These show consistency with the potential of
Eq.~(\ref{total_potential}).
\acknowledgments
We thank Oliver Jahn for providing
us with the detailed notes of his calculation of
the baryonic potential. We also thank O.~Jahn and Philippe de Forcrand
for enlightening discussions. 
M.P.\ gratefully acknowledges financial support from the
Alexander~von~Humboldt Foundation and from INFN. The University
of Regensburg hosts the Collaborative Research Center
SFB/TR 55 ``Hadronenphysik mit Gitter-QCD''.
\appendix
\section{Conformal
mapping of a world sheet to a rectangle}\label{AppA}
Here we provide the ingredients for the calculation of the
fluctuations at the junction.
We follow Ref.~\cite{Jahn:2003uz}, conformally mapping the
blade (see Fig.~\ref{fig:blades_and_one_blade}) to a rectangle.
The minimal-area solution for a fixed position of the junction,
$\bmath{\xi}_{min,a}(t,s)$, is harmonic and satisfies the boundary
conditions Eq.~(\ref{boundarycondition_xi}):
\begin{equation}
\triangle \bmath{\xi}_{min,a} = 0\,, \qquad \bmath{\xi}_{min,a}(t,L_a+\bmath{ e}_a \cdot\bmath{ \varphi}(t))=\bmath{ \varphi}_{\perp a}(t)\,.
\end{equation}
The determinant in Eq.~(\ref{partitionfct_string}) is computed with
Dirichlet boundary conditions on the domain $\Gamma_a=\{(t,s)|0\leq
s\leq L_a+\bmath{ e}_a \cdot \bmath{ \varphi}(t)\}$. In terms of the Fourier
components $\bmath{ \varphi}_w$ of $\bmath{ \varphi}(t)$,
$\bmath{\xi}_{\min, a }$ is
given by
\begin{equation}
  \bmath{\xi}_{min,a}=\frac{1}{\sqrt T}\sum_w \bmath{ \varphi}_{w,\perp a}
  \frac{\sinh(ws)}{\sinh(w L_a)} e^{iwt} + {\cal O}
  (\varphi^2)\:,
\end{equation}
where $w={2\pi n}/{T}$.
The integral in Eq.~(\ref{partitionfct_string}), which represents the
change in the minimal area due to
the transverse fluctuations $\bmath{ \varphi}_{\perp a}$, can now be calculated:
\begin{eqnarray}
  \int_{\Gamma_a}\!\!d^2\!{\rm \zeta}\,\sum_{i} \frac{\partial
  \bmath{ \xi}_{\min,a}}{\partial {\rm \zeta}_i} \cdot
\frac{\partial \bmath{\xi}_{\min,a}}{\partial {\rm \zeta}_i} &=& \sum_w w \coth(w L_a)|\bmath{ \varphi}_{w,\perp a}|^2 \nonumber \\
  && + {\cal O} (\varphi^3)\:.
\end{eqnarray}
The determinant in Eq.~(\ref{partitionfct_string}) is obtained
by mapping the domain $\Gamma_a$ conformally to a rectangle $L'_a
\times T$. Note that the conformal map $f_a(z)=z+\sum_w c_{w a}
e^{wz}$ has to be complex differentiable. Its coefficients $c_{w a}$
are fixed by the constraints:
\begin{eqnarray}
  f_a(i\mathbb{R}) &=& i \mathbb {R}\:, \nonumber\\
  f_a(L'_a + i t) &=& L_a + \bmath{ e}_a \cdot \bmath{ \varphi} (t) + i t + {\cal
  O}(\varphi^2)\:.
\end{eqnarray}
One easily sees that $L'_a = L_a + \frac{1}{\sqrt{T}}\bmath{ e}_a \cdot
\bmath{ \varphi}_0 $. To leading order in
$\bmath{ \varphi}$ the conformal map is then
given by
\begin{equation}\label{conformal map}
f_a(z)=z+\frac{1}{\sqrt T} \sum_{w\neq 0} \frac{\bmath{ e}_a \cdot \bmath{
\varphi}_w}{\sinh(w L_a)} e^{wz} + {\cal O}(\varphi^2)\:.
\end{equation}
This conformal mapping
changes the Laplacian by
a scalar factor:
\begin{equation}
 \triangle_{\Gamma_a}=e^{2\rho_a(z)} \triangle_{L'_a\times T}\:,\quad
\rho_a(z)=- \frac 1 2 \ln |\partial_z f_a|^2\:.
\end{equation}
The variation of the determinant of the Laplacian with respect to a
holomorphic mapping of $\Gamma$ onto some other region $\tilde
\Gamma$ via the function $f(z)$ can be calculated by means of the
Alvarez-Polyakov formula (see e.g.\ Ref.~\cite{Luscher:1980fr}): 
\begin{eqnarray}
  &&\ln \frac{\det (-\triangle_\Gamma)}{ \det
    (-\triangle_{\tilde\Gamma})}=
  \frac{1}{12\pi} \int_{\partial\Gamma}\!d\tau\,
    \frac{\epsilon_{ij}z'^i
    z''^j}{z'^2}\ln |\partial_z f|^2 \nonumber \\
  &&+\frac{1}{12\pi}\int_{\Gamma}\!d^2\!z\,
    \partial_z \ln |\partial_z f|^2 \partial_{\bar{z}}\,\ln |\partial_z f|^2\:.
\end{eqnarray}
Here $z(\tau)$ is an arbitrary parametrization of $\partial
\Gamma$ and $z'=dz/d\tau$.
In our case the first integral above
vanishes and thus, from the conformal map
Eq.~(\ref{conformal map}), we obtain to leading order
\begin{eqnarray}
&&\int_{L'_a \times T}\!d^2\!z\,
  \partial_z \ln |\partial_z f_a|^2\, \partial_{\bar{z}} \ln |\partial_z
  f_a|^2 \nonumber \\
  &=& \sum_w w^3 |\bmath{ e}_a \cdot \bmath{
  \varphi}_w|^2 \coth(w L_a)+{\cal O}(\varphi^3)\:,
\end{eqnarray}
where we used the fact that the Fourier coefficients satisfy
$\bmath{ \varphi}_{-w}=\bmath{ \varphi}_w^*$.
In Ref.~\cite{Luscher:1980fr} the rectangle with periodic boundary
conditions in time is further conformally mapped onto a circle resulting
in:
\be
\det(-\triangle_{L'_a\times T})=\eta^2\!\left(
\frac{iT}{2L'_a}\right)\:,
\en
where $\eta(\tau)$ denotes the Dedekind $\eta$ function. Collecting
the above results,
we obtain for the determinant of the Laplacian with respect to the blade $a$:
\begin{eqnarray}
\label{barypotgaussian}
&&\!\!\!\!\!\!\!\!\!\!\!\det(-\triangle_{\Gamma_a}) = \eta^2\!\left( \frac{iT}{2L'_a}\right) \nonumber \\
&& \!\!\!\!\!\!\!\!\!\!\!\times \exp\left(- \frac{1}{12\pi}\sum_w w^3 \coth (w L_a)|\bmath{ e}_a \cdot \bmath{
  \varphi}_w|^2\right)\:.
\end{eqnarray}

\section{Calculation of the width of the junction
$\langle{\bmath{ \varphi}}^2\rangle$}\label{Appendix_mu2}
In this appendix we provide more steps for the calculation of
Eqs.~(\ref{iperp}) and (\ref{eq_phi_parallel}). The calculation is
carried out for $n$ quarks located in a plane. This configuration
with only one junction might not be stable for more than three
quarks~\cite{Gliozzi:2005en}.
However, the result for the orthogonal contribution can easily be
generalized to configurations of $n$ quarks, distributed
in a $(n-1)$-dimensional hyperplane.

The thickness of the string at the junction can be calculated taking
the expectation value of ${\bmath{ \varphi}}^2$ [see
Eq.~(\ref{generalaction})]:
\be
  \langle{\bmath{ \varphi}}^2\rangle=
\frac{\int\! {\EuScript D }\bmath{ \varphi}\,{\bmath{ \varphi}}^2
  e^{-S}}{\int\! {\EuScript D}\bmath{ \varphi}\,e^{-S}}\:.
\en
To do this, we have to consider integrals
\begin{eqnarray}
\label{App_Int1}
  && \!\!\! \int\!{\EuScript D}\bmath{ \varphi}\,\exp
\left[-\frac{m}{2}\int_0^T\!dt\,|\dot{\bmath{
  \varphi}}|^2 \right.\nonumber\\
&&+\sum_{a=1}^n
\left(-\frac{\sigma}{2}\int\!d^2\!{\rm \zeta}\,\sum_i
\frac{\partial
  \bmath{ \xi}_{\min,a}}{\partial {\rm \zeta}_i}
\cdot \frac{\partial \bmath{\xi}_{\min,a}}
{\partial {\rm \zeta}_i} \right. \nonumber\\
&&\left.\left.+ \frac{D-2}{24\pi}\sum_w w^3 \coth (w L_a)|\bmath{ e}_a \cdot \bmath{
  \varphi}_w|^2\right)\right]\:.
\end{eqnarray}
We can replace the integral in the first term above
by a sum over Fourier components:
\be
\int_0^T\! dt\, |\dot{\bmath{ \varphi}}|^2
  =\sum_w w^2 |\bmath{ \varphi}_w|^2\:.\en
We denote the plane that is spanned by the
spatial unit vectors of the $n$ strings
$\bmath{ e}_a=(e_{a,x},e_{a,y},0,\ldots)$ as the $x$-$y$ plane.
These $n$ unit
vectors obey the relation $\sum_a {\bmath{ e}}_a =\bmath{ 0}$.
The $x$-$y$ components of
$\bmath{ \varphi}$ (or any other vectors) carry the
superscript ``$\parallel$''. We obtain
\begin{eqnarray}
 && \!\!\!   \sum_a\coth (w L_a)|\bmath{ e}_a \cdot \bmath{ \varphi}_w^\parallel|^2 \nonumber \\
 && \!\!\!= |\varphi_{w,x}^\parallel|^2 \sum_a e_{a,x}^2 \coth(w L_a) +  |\varphi_{w,y}^\parallel|^2 \sum_a e_{a,y}^2 \coth(w L_a)\nonumber\\
 && + 2 \left( {\mbox{Re}} (\varphi_{w,x}^\parallel) {\mbox{Re}}(\varphi_{w,y}^\parallel)+{\mbox{Im}}( \varphi_{w,x}^\parallel){\mbox{Im}}(\varphi_{w,y}^\parallel)\right) \nonumber \\
 && \times \sum_a e_{a,x} e_{a,y} \coth(w L_a)\:.
\end{eqnarray}
We define $A_x,A_y$ and $A_{\mbox{\tiny{Re}}}$ as
\begin{align}
  &A_x = \left( \frac{\sigma} {2} w + \frac{(D-2)w^3}{24\pi} \right)
            \left[
            \sum_a e_{a,x}^2 \coth(w L_a)
            \right]\:,\nonumber\\
  &A_y = \left( \frac{\sigma} {2} w + \frac{(D-2)w^3}{24\pi}\right)
            \left[
            \sum_a e_{a,y}^2 \coth(w L_a)
            \right]\:,\\
 &A_{\mbox{\tiny{Re}}} = \left( \frac{\sigma} {2} w + \frac{(D-2)w^3}{24\pi} \right)
            \left[
            \sum_a e_{a,x} e_{a,y} \coth(w L_a)
            \right]\:.\nonumber
\end{align}
Note that $A_x + A_y$ as well as the combination \mbox{$A_xA_y-A_{\mbox{\tiny{Re}}}^2$}
are invariant under rotations within the $x$-$y$ plane. Since the
angles between the $n$ strings are equal, we can parametrize the unit
vectors by
\begin{equation}
\bmath{e}_a=(\cos({2\pi}a/n),\sin({2\pi}a/n),0,\ldots)\:.
\end{equation}
Thus we obtain
\begin{eqnarray}
A_xA_y-A_{\mbox{\tiny{Re}}}^2 &=& \left( \frac{\sigma} {2} w +
\frac{(D-2)w^3}{24\pi} \right)^2 \sum_{a<b}
\underbrace{\sin^2\left(2\pi\frac{(a-b)}n\right)}_{\alpha_{ab}} \nonumber \\ 
&& \times \coth(wL_a)\coth(w L_b)\:.
\end{eqnarray}
In the case of $n=3$, the geometrical
coefficients are $\alpha_{ab}=3/4$. This results in the simplification:
$A_x A_y -A_{\mbox{\tiny{Re}}}^2=\frac 3 4 \left(
\frac{\sigma} {2} w + \frac{(D-2)w^3}{24\pi} \right)^2\sum_{a <
b}\coth(w L_a)\coth(w L_b)$.

We can split the integral over $\bmath{ \varphi}$ in Eq.~(\ref{App_Int1})
using $|\bmath{ \varphi}_{w,\perp a}|^2=|\bmath{ \varphi}_{w}|^2-|\bmath{ \varphi}_w\cdot\bmath{
e}_a|^2$ into parts that are parallel and perpendicular to the plane
of the quarks:
\begin{widetext}
\begin{eqnarray}
&&\!\!\!\!\int\!{\EuScript D}\bmath{ \varphi}\exp\left[-\frac{m}{2}\!
\int\! dt\,|\dot{\bmath{
  \varphi}}|^2\!+\sum_{a=1}^3\!\left(\!-\frac{\sigma}{2}\!
\int\! |\partial \bmath{\xi}_{\min,a}|^2+\!\frac{D-2}{24\pi}
\sum_w\!\! w^3\! \coth (w L_a)|\bmath{ e}_a \cdot \bmath{
  \varphi}_w|^2\!\right)\right]\nonumber\\
&=&\int\!{\EuScript D}\bmath{ \varphi}^\perp
\exp\left[
-\frac{1}{2}\sum_w
\left(m w^2+ \sigma w \sum_a\coth(w L_a)\right)
            |\bmath{ \varphi}_{w}^\perp|^2\right]\nonumber\\
  &&\times\:\int\!{\EuScript D}\bmath{ \varphi}^\parallel
\exp\left\{\sum_w\Big[-\frac{1}{2}\left(m w^2+
\sigma w \sum_a\coth(w L_a) \right)
|\bmath{ \varphi}_{w}^\parallel|^2 +|\varphi_{w,x}^\parallel|^2 A_x + |\varphi_{w,y}^\parallel|^2 A_y
 + 2\left( {\mbox{Re}}(\varphi_{w,x}^\parallel)
{\mbox{Re}}(\varphi_{w,y}^\parallel) \right.\right. \nonumber \\
 && \left. + {\mbox{Im}} (\varphi_{w,x}^\parallel) {\mbox{Im}} (\varphi_{w,y}^\parallel)\right) A_{\mbox{\tiny{Re}}} \Big] \Bigg\} \:.
\end{eqnarray}
\end{widetext}
Here, $\bmath{ \varphi}_w^\perp$ are the $D-3$ components
of  $\bmath{ \varphi}_w$ that are
perpendicular to the plane spanned by the quarks.
We abbreviate the first functional integral above
as $I_1$ and the second as $I_2$.
The solutions of these Gaussian integrals read
\begin{eqnarray}
  I_1&=&
  \left(
  \prod_{w>0} \frac{\pi}{m w^2
            +{\sigma} w \sum_a\coth(w
            L_a)}
  \right)^{D-3}\:,\\
  I_2 &=&   \prod_{w>0} \frac{\pi^2}{A_1 A_2 -4A_{\mbox{\tiny{Re}}}^2}\:,\label{res_I2}
\end{eqnarray}
where we defined
\begin{eqnarray}
    C_w
&=&
    m w^2+ \sigma w \sum_a\coth(w L_a)\:,
\nonumber\\\label{App_def_A1_A2}
    A_1
&=&
    C_w - 2A_x\:,
\\
    A_2
&=&
    C_w - 2A_y\:.
\nonumber
\end{eqnarray}
We are interested in the expectation value
\begin{equation}
  \langle{\bmath{ \varphi}}^2\rangle=\langle{\bmath{ \varphi}}^{\perp 2}\rangle+\langle{\bmath{ \varphi}}^{\parallel 2}\rangle=\frac{I_\perp}{I_1}+\frac{I_\parallel}{I_2}\:,
\end{equation}
where
\begin{eqnarray}
  &&\!\!\!\!\!\!\!\!\!\!I_\perp =  \!\!\int\!\! {\EuScript D}\bmath{ \varphi}^\perp{\bmath{ \varphi}}^{\perp 2}\exp\Bigg\{-\frac12\sum_w \Big[ mw^2 + \sigma w \sum_a\coth(wL_a)\Big]\nonumber\\
  && \;\;\times |\bmath{ \varphi}_w^\perp|^2\Bigg\}\:,\\
  &&\!\!\!\!\!\!\!\!\!\! I_\parallel = \!\!\int\!\! {\EuScript D}\bmath{ \varphi}^\parallel {\bmath{ \varphi}}^{\parallel2}\exp\Bigg\{\sum_w\Big[-\frac{1}{2}\Big(m w^2 + \sigma w \sum_a\coth(w L_a) \Big)\nonumber\\
  && \;\; \times |\bmath{ \varphi}_{w}^\parallel|^2 +|\varphi_{w,x}^\parallel|^2 A_x + |\varphi_{w,y}^\parallel|^2 A_y + |\varphi_{w,y}^\parallel|^2 A_y \nonumber \\
  && \!\!\!\!\!\!\!\!\!\!\!\!+ 2\left( {\mbox{Re}}(\varphi_{w,x}^\parallel) {\mbox{Re}}(\varphi_{w,y}^\parallel) +{\mbox{Im}}(\varphi_{w,x}^\parallel) {\mbox{Im}}(\varphi_{w,y}^\parallel)\right) A_{\mbox{\tiny{Re}}}\Big]\Bigg\}\:.\label{App_Iparallel}
\end{eqnarray}
Let us recall that $\bmath{ \varphi}^\perp$ is $(D-3)$-dimensional.
Performing the first integral yields
\be
  I_\perp
  =
        (D-3) \frac 2 T
        \left( \prod_{w'>0} \frac{\pi}{C_{w'}}\right)^{D-3}\sum_{w>0}
        \frac{1}{C_w}\:,
\en
so that
\begin{equation}
  \langle{\bmath{ \varphi}}^{\perp 2}\rangle=\frac{I_\perp}{I_1} =  \frac 2 T \sum_{w>0}         \frac{(D-3)}{mw^2+{\sigma} w \sum_a\coth(wL_a)}\label{tildeI1_I1}\:.
\end{equation}
With $w = 2 \pi n/T$, in the limit of large $T$ we obtain
\eq{App_mu_perp}
  \langle{\bmath{ \varphi}}^{\perp 2}\rangle =  \frac 1 \pi \int_0^\infty\!{dw}\,\frac{D-3}{mw^2+{\sigma} w \sum_a\coth(wL_a)}\:.
\en

The parallel contribution to the width of the junction is calculated
analogously. We perform the integral $I_\parallel$
[Eq.~(\ref{App_Iparallel})]. In terms of $A_1$ and $A_2$ defined in
Eq.~(\ref{App_def_A1_A2}), one is left with
\be
  I_\parallel              = \frac {2}{T}
                 \left(\prod_{w>0} \frac {\pi^2}{A_1 A_2-4A_{\mbox{\tiny{Re}}}^2}\right)
                 \sum_{w>0}
                 \frac {A_1 + A_2}{A_1 A_2-4A_{\mbox{\tiny{Re}}}^2}\:.
\en
Therefore, combining $I_\parallel$ with Eq.~(\ref{res_I2}) yields
\begin{widetext}
\begin{equation}
  \langle{\bmath{ \varphi}}^{\parallel 2}\rangle = \frac{I_\parallel}{I_2}
      = \frac {2}{T}
                 \sum_{w >0}
                 \frac {A_1 + A_2}{A_1 A_2-4A_{\mbox{\tiny{Re}}}^2}
                 = \frac 4 3 \frac {1}{\sigma\pi}
                \int_0^\infty dw \frac 1 {w}
                 \frac {\tilde{m} w+(1-a w^2) C_1}
                        {\tilde{m}^2w^2 +2\tilde{m}w(1-aw^2) C_1-4aw^2 \left(C_1\right)^2+ \frac 4 3 (1+a w^2)^2
                        C_2}\:,
\label{ipar}
\end{equation}
\end{widetext}
where
\begin{eqnarray}
\tilde{m}&=&\frac 2 3 \frac m \sigma\:,\quad a=\frac{D-2}{12\pi\sigma}\:,
\\
 C_1 &=& \frac 1 3 \sum_a \coth(w L_a)\:,
\\
 C_2 &=& \frac 1 3  \sum_{a< b}\alpha_{ab}\coth(w L_a)\coth(w L_b)\:.
\end{eqnarray}
For an equilateral baryon we have $\sum_{a<
b}\alpha_{ab}=\frac 9 4$ and $L=L_1=L_2=L_3$. Thus, for this special
case, we obtain
\eq{mupara_bary}
\langle{\bmath{ \varphi}}^{\parallel 2}_{qqq,\triangle}\rangle = \frac {4}{3\sigma\pi} \int_0^\infty \frac {dw}{w^2\tilde{m}/L + (w- w^3 a/L^2) \coth(w) }\:.
\en


\end{document}